\documentclass[11pt]{iopart}

\usepackage{iopams}
\usepackage{setstack}
\usepackage{hyperref} 
\usepackage{bm}
\usepackage{longtable}
\usepackage{graphicx}
\usepackage[margin=2cm]{geometry}
\usepackage{pdflscape}
\usepackage{cite}
\usepackage{subcaption}

\newcommand{\dd}{\mathrm{d}}

\hypersetup{ colorlinks=true, linktoc=all, linkcolor=blue, }
\begin{document}

\title{Emergent Cosmological Expansion in Scalar-Tensor Theories of Gravity}

\author{Chad Briddon$^1$, Timothy Clifton$^1$ and Pierre Fleury$^{2, 3}$}
\address{$^1$Department of Physics \& Astronomy, Queen Mary University of London, UK.}
\address{$^2$Laboratoire Univers et Particules de Montpellier (LUPM), 
CNRS \& Universit\'{e} de Montpellier (UMR-5299),
Place Eug\`{e}ne Bataillon, F-34095 Montpellier Cedex 05, France.}
\address{$^3$Universit\'{e} Paris-Saclay, CNRS, CEA, Institut de physique th\'{e}orique, 91191, Gif-sur-Yvette, France.}

\begin{abstract}

We consider the emergence of large-scale cosmological expansion in scalar-tensor theories of gravity. This is achieved by modelling sub-horizon regions of space-time as weak-field expansions around Minkowski space, and then subsequently joining many such regions together to create a statistically homogeneous and isotropic cosmology. We find that when the scalar field can be treated perturbatively, the cosmological behaviour that emerges is well modelled by the Friedmann solutions of the theory. When non-perturbative screening mechanisms occur this result no longer holds, and in the case of scalar fields subject to the chameleon mechanism we find significant deviations from the expected Friedmann behaviour. In particular, the screened mass no longer contributes to the Klein-Gordon equation, suppressing deviations from general relativistic behaviour.

\end{abstract}

\section{Introduction}\label{sec:introduction}

The large-scale properties of the Universe are most usually modelled using a homogeneous and isotropic Friedmann-Lema\^{i}tre-Robertson-Walker (FLRW) cosmological model, with linear perturbation theory describing structures within it \cite{malik}. While this approach is highly successful at interpreting a wide range of observable phenomena, it does come with certain drawbacks. In particular, it does not allow one to model non-linear structures without appealing to Newtonian gravity \cite{peebles}, and it does not allow the effects of non-linear structure formation to back-react onto the large-scale cosmology \cite{buchert}. Addressing these points is the purpose of the post-Newtonian cosmologies created in Refs. \cite{noav, Sanghai:2015wia, Sanghai:2016ucv}.

In the post-Newtonian approach to cosmological modelling one takes small (sub-horizon) regions of space-time to be approximated by a post-Newtonian expansion of Minkowski space-time, and then joins these regions together using appropriate junction conditions in order to obtain a cosmological model. This approach very naturally allows for non-linear structures to be included in a self-consistent way, as almost all structures are readily modelled using post-Newtonian expansions (with the notable exceptions of black holes and neutron stars) \cite{poisson}. These types of models also allow the cosmological expansion to emerge as a natural consequence of the gravitational fields of the bodies that exist within them, thereby allowing the back-reaction of structure on the large-scale properties of the Universe to emerge without having to pre-suppose the validity of modelling averages using homogeneous and isotropic perfect fluids \cite{Fleury:2016tsz}.

This approach has been well-studied in the context of Einstein's theory of General Relativity (GR), where it has been shown that reflection symmetric boundary conditions between neighbouring regions of post-Newtonian expanded space-time give dust-dominated FLRW behaviour at leading-order \cite{noav}, and radiation-like terms at next-to-leading order \cite{Sanghai:2015wia, Sanghai:2016ucv}. Within GR, the behaviour of perturbations on very large scales has also been investigated \cite{gold1, gold2, gal1}, as has the gauge problem \cite{gold3}, the consequences for the bispectra of the matter distributions \cite{gal2, gal3}, and the resultant Hubble diagrams \cite{pierre}. It has even been applied to $f(R)$ theories \cite{dun1, dun2}, and to parameterizations of theories of gravity \cite{ppnc1, ppnc2, ppnc3, ppnc4}. In the present work we wish to extend the application of this treatment to scalar-tensor theories of gravity, which are some of the most widely studied in the literature on alternatives to Einstein's theory \cite{review1, review2}.

In Section \ref{sec:theory} of this paper we introduce the class of scalar-tensor theories we wish to study, along with the Friedmann and scalar field equations that govern homogeneous and isotropic space-times within them. In Section \ref{sec:PN_cosmo} we then introduce the post-Newtonian expansions, and the way that we will build cosmological models by applying them. This includes, in particular, a discussion of the conditions on the geometry and the scalar field at the junction between neighbouring regions of post-Newtonian expanded Minkowski space. Section \ref{sec:gen} then gives a derivation of the large-scale properties of a lattice construction built from many such regions of space, assuming that the metric is post-Newtonian expanded, but making no restrictions on the scalar field. This is then specialized in Section \ref{sec:pert} to the case where the scalar field can be treated as being perturbatively expanded. We show that in this case the Friedmann solutions of the theory are precisely recovered. 

We then consider non-perturbative screening mechanisms in Section \ref{sec:screen}, where we show using the Chameleon mechanism \cite{cham} that screened mass is removed from the scalar field equation obeyed by the large-scale cosmology. The regions of space within a screened body also do not contribute to the emergent Friedmann equations in the usual way, when one tallies up the total energy density in the scalar field, as the scalar is forced to a fixed value defined by the density of mass in its local environment. This shows that non-linear screening mechanisms can have a strong back-reaction effect on the large-scale cosmological expansion, and that the Friedmann equations of a theory may not be valid when screening occurs. We conclude in Section \ref{sec:disc}.

\section{Scalar-tensor theories of gravity} \label{sec:theory}

The gravitational theories we wish to consider in this work are mediated by a metric tensor field $g_{\mu \nu}$ and a scalar field $\phi$. They are defined by the following action \cite{bergman, nordvedt, wagoner}:
\begin{equation} \label{action}
\hspace{-2.3cm}
S[g_{\mu\nu}, \phi, \psi]
= \frac{1}{16\pi G} \int\dd^4 x \sqrt{-g} \; (R-2\Lambda)
	- \int \dd^4 x \sqrt{-g} \, \left[ \frac{1}{2}\,\partial_{\mu} \phi \partial^{\mu}\phi  + V(\phi) \right]
	+ S_{\rm m}[\psi; C^2(\phi) g_{\mu\nu}] \ ,
\end{equation}
where $R$ is the Ricci scalar constructed from $g_{\mu \nu}$, $\Lambda$ is the cosmological constant, and $S_{\rm m}$ denotes the action of all matter fields $\psi$. {  The independent functions $V(\phi)$ and $C(\phi)$ describe the self-interaction potential of the scalar field and a universal coupling between the scalar field and the matter fields\footnote{{  In principle one could allow different couplings between the scalar field and different types of matter fields, but this would result in a violation of the Weak Equivalence Principle (WEP), which we do not wish to consider here (see Ref. \cite{micro} for current constraints on WEP violations from the MICROSCOPE mission).}}, respectively}. The choice of writing our theories of gravity in this way shows that our action has been constructed in the ``Einstein frame'' \cite{dicke}. The coupling $C(\phi)$ could be removed by defining a new metric $\tilde{g}_{\mu \nu} \equiv C^2(\phi) g_{\mu \nu}$, at the expense of complicating the coupling between $\phi$ and $g_{\mu \nu}$ in the rest of the action, which would result in formulating physics in the ``Jordan frame'' \cite{dicke}. We choose not to do this here, as it would complicate the form of the corresponding field equations, and because the chameleon mechanism that we study is usually defined in the Einstein frame \cite{cham}. We will, however, refer back to the Jordan frame when stating our results in later sections.

Extremizing the action (\ref{action}) with respect to the metric $g_{\mu \nu}$ results in the following field equations:
\begin{eqnarray}
\label{eq:field}
&&R_{\mu\nu} - \frac{1}{2}\,R\, g_{\mu\nu} + \Lambda\,g_{\mu\nu}
= 8\pi G \left[ |C(\phi)| T_{ \mu\nu}^{\rm m} + T_{\mu\nu}^\phi \right] 
\end{eqnarray}
where we have assumed $C(\phi)>0$, and where the energy-momentum tensors of scalar and matter fields are defined by
\begin{eqnarray} \nonumber
\hspace{-1cm}
 T^{\phi}_{ \mu\nu} &\equiv& \partial_{\mu} \phi \partial_{\nu} \phi - g_{\mu \nu} \left[ {  \frac{1}{2}} \partial^{\sigma} \phi \partial_{\sigma} \phi + V(\phi) \right]\, , \qquad
T^{{\rm m}}_{\mu\nu} \equiv -\frac{2}{C(\phi)\sqrt{-g}} \frac{\delta S}{\delta g^{\mu\nu}} \ .
\end{eqnarray}
Specified in this way, the quantity $\rho \equiv T_{\mu \nu}^{\rm m} u^{\mu} u^{\nu}$ satisfies the usual conservation equation for energy (though one may note that the gravitational energy in the Einstein frame is then $|C(\phi)| \, \rho$). 

Extremizing (\ref{action}) with respect to the scalar field $\phi$ gives
\begin{eqnarray}
\label{eq:scalar}
&&\square\phi = V'(\phi) + C'(\phi) T^{\rm m} \, ,
\end{eqnarray}
where $T^{\rm m} = g^{\mu\nu} T_{\mu\nu}^{\rm m}$ is the trace of the energy-momentum tensor of matter, {  and where the primes denote derivatives with respect to $\phi$}. In this theory the equation of motion of a spin-free test particle with four-velocity $u^\mu$ is given by
\begin{eqnarray}
\label{eq:motion}
&&u^\nu \nabla_\nu \left[ |C(\phi)| u^\mu \right] = - |C(\phi)|' \partial^\mu \phi \ .
\end{eqnarray}
We can see from these equations that $C(\phi)$ controls the degree to which matter gravitates, and that for non-constant $C(\phi)$ we have both a source in the scalar field equation (\ref{eq:scalar}) and a fifth-force acting in the equation of motion of test particles (\ref{eq:motion}).

Within this class of theories, the Friedmann equations governing the expansion of homogeneous and isotropic cosmologies can be written as
\begin{equation} \label{frw1}
\left(\frac{\dot{a}}{a} \right)^2 = \frac{8 \pi G C}{3} \rho - \frac{\kappa}{a^2} +\frac{\Lambda}{3} +\frac{8\pi G}{3} \left( \frac{1}{2} \dot{\phi}^2 +V \right) \, ,
\end{equation}
and
\begin{equation} \label{frw2}
\hspace{0.8cm} \frac{\ddot{a}}{a} = - \frac{4\pi G C}{3} (\rho + 3 p) + \frac{\Lambda}{3} - \frac{8 \pi G}{3} \left( \dot{\phi}^2-V \right) \, ,
\end{equation}
where $a=a(t)$ is the scale factor of the cosmology, and $\kappa$ is the unit curvature scalar of spatial sections. The energy density $\rho$ is defined as above, and $p \equiv \frac{1}{3} T^{\rm m}_{\mu \nu} (g^{\mu \nu} + u^{\mu} u^{\nu})$ is the isotropic pressure of the matter fields. The corresponding equation for the scalar field is
\begin{equation} \label{frw3}
\ddot{\phi} +3 \frac{\dot{a}}{a} \dot{\phi} + V' +C' (\rho-3p) =0 \, ,
\end{equation}
where throughout we have taken $\phi=\phi(t)$, consistent with the symmetries of the space-time. Equations (\ref{frw1})--(\ref{frw3}) are widely taken to represent the large-scale behaviour of cosmologies in these theories, up to the addition of small perturbations in order to account for inhomogeneities. It is our goal in what follows to test this assumption in the presence of non-linear structures.

\section{Post-Newtonian cosmology} \label{sec:PN_cosmo}

\begin{figure}
\center
\includegraphics[width=11cm]{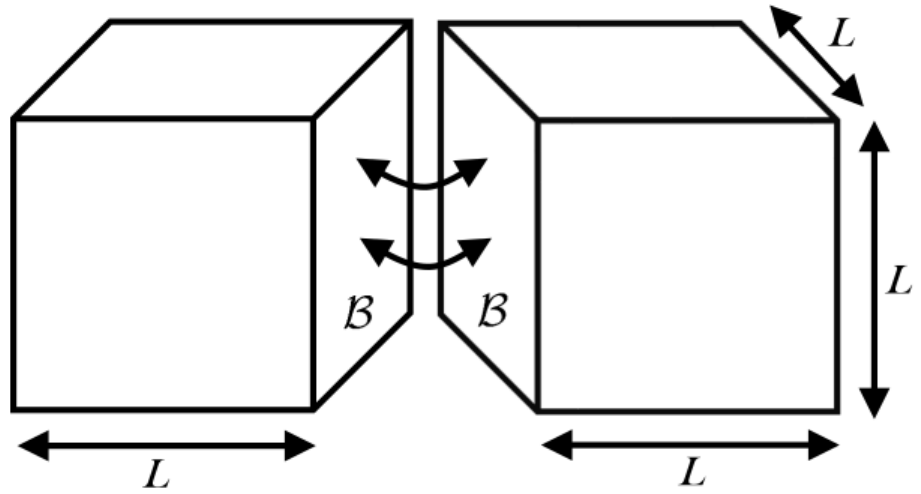}
\caption{Two cubic cells of side length $L$ being joined together at a common boundary $\mathcal{B}$, whose union constitutes a solution to the field equations of the theory (\ref{action}), with no boundary layer, if Eq.~(\ref{eq:junction}) is satisfied.}  \label{cubes}
\end{figure}

We wish to pursue the post-Newtonian approach to cosmological modelling, as expounded in Ref. \cite{Sanghai:2015wia}. The goal here is to link the large-scale cosmological expansion to the weak gravitational fields of objects that exist within the space-time, through a patchwork structure in which many sub-horizon-sized cells are joined together to create a global cosmological space-time. This concept is illustrated in Fig. \ref{cubes}, where two cubic cells are being joined together at a common boundary $\mathcal{B}$, in order to form a larger spatial volume that is also a solution of the theory.

In order for the union of these two cells to constitute a solution of the field equations (\ref{eq:field}) and (\ref{eq:scalar}) we require the following junction conditions to be satisfied \cite{junction}:
\begin{equation} \label{eq:junction}
[g_{\mu \nu} - n_{\mu}n_{\nu}]^+_-=[K_{\mu \nu}]^+_-=0 \qquad {\rm and} \qquad [\phi]^+_- = n^{\mu} [ \partial_{\mu} \phi ]^+_- = 0 \, ,
\end{equation}
where we have assumed that there is no surface layer of energy-momentum on the boundary, {  and that the topology of the boundary is trivial}, as appropriate for a simple cosmological model. The notation $[\dots ]^+_-$ means the difference in the quantity contained in brackets when evaluated on either side of the boundary,  and where it is assumed that a common set of coordinates is constructed on the boundary when evaluating tensorial objects. The vector components $n^{\mu}$, in these expressions, correspond to the space-like unit normal to the boundary, and $K_{\mu \nu}$ is the extrinsic curvature of the boundary.

The first set of equations in (\ref{eq:junction}) give the conditions that the first and second fundamental forms of the boundary are the same on either side, and the second set tell us the scalar field and its normal derivative must be continuous. These conditions are required for the Ricci scalar $R$ and the d'Alembertian $\square \phi$ to remain finite at the boundary, such that no divergences occur in the field equations (\ref{eq:field}) and (\ref{eq:scalar}). With these conditions satisfied, we can treat the geometry of space-time within each cell as being described by a post-Newtonian expansion \cite{will}, while building a global cosmological space-time. This approach is expected to be valid as long as the size of each cell is small compared to the horizon, such that the motion of its boundary satisfies $v/c \ll 1$ (i.e. that it is sub-horizon-sized), and as long as we are not trying to model the geometry of space-time in the vicinity of compact objects such as black holes and neutron stars. In what follows, we will assume for simplicity that the boundary is reflection symmetric, such that $K_{\mu \nu} = 0 = n^{\mu}  \partial_{\mu} \phi$ at $\mathcal{B}$.

The post-Newtonian expansion is a weak-field and slow-motion expansion about Minkowski space, meaning that at leading-order the metric is described by \cite{will}
\begin{equation} \label{pnmetric}
\dd s^2 = -(1+2\Phi) \dd t^2 + (1-2\Psi) \delta_{ij} \dd x^i \dd x^j + \mathcal{O}(\epsilon^3) \ ,
\end{equation}
where $\Phi, \Psi \sim \epsilon^2$ are the (post-)Newtonian gravitational potentials, and where the smallness parameter $\epsilon \sim v/c \ll 1$. Provided that the length scales under consideration are sufficiently small, the slow-motion requirement means that time derivatives of objects are much smaller than their spatial derivatives, such that
\begin{equation}
\frac{\partial}{\partial t} \sim \epsilon \, \frac{\partial}{\partial x^i} \ .
\end{equation}
We will also make the standard assumption that matter in the Universe is well-described as dust, so that $p\sim \epsilon^2 \rho \sim \epsilon^4$. Applying this type of expansion allows us to describe the near-zone gravitational fields of almost all structures in the Universe, even deep into the non-linear regime \cite{poisson}.

The consequences of the first and second fundamental forms matching smoothly at the boundary has already been investigated in Ref. \cite{Sanghai:2015wia}, and plays out similarly here. The conditions that the scalar field $\phi$ should be continuous and have a $C^1$ smooth normal derivative is, however, new. In order to understand the consequences of this latter condition under reflection symmetry, let us write
\begin{equation}\label{eq:boundary_condition}
0 = n^\mu \partial_\mu \phi = n^t \dot{\phi} + \vec{n} \cdot \vec{\nabla} \phi \ ,
\end{equation}
where arrows indicate 3-vectors. Now, the space-like normal $n^{\mu}$ has (by definition) no time-like component in the frame of reference of an observer that is comoving with the boundary. The difference between quantities measured by this observer, and a second observer who is stationary in the coordinates $\{t, x^i\}$ of (\ref{pnmetric}), is then given by a Lorentz transformation, which tells us that $n^t = v + \mathcal{O}(v^3)$ and $\vert \vec{n} \vert = 1 + \mathcal{O}(v^2)$. We therefore have that on the boundary, at leading order,
\begin{equation} \label{dphi}
\vec{n} \cdot \vec{\nabla} \phi = -v \, \dot{\phi} = - \frac{1}{2} \, H \, L \, \dot{\phi} \, ,
\end{equation}
where $H$ is the Hubble rate of expansion of the cosmology, $L$ is the width of one cell, and where we have assumed that the global cosmology is expanding at equal rates in every direction.

\begin{figure} 
\center
\includegraphics[width=9cm]{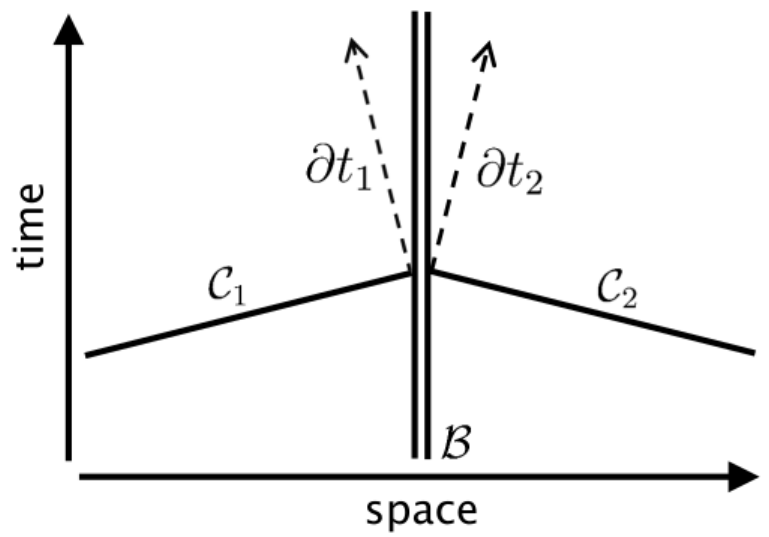}
\caption{The spatial sections of constant $t$ from two neighbouring cells, $\mathcal{C}_1$ and $\mathcal{C}_2$, meeting at their common boundary $\mathcal{B}$. The surfaces do not overlap, and are not orthogonal to the boundary.} \vspace{-0.37cm}\label{boundary}
\end{figure}

The geometric picture of what is happening at the boundary is illustrated in Fig. \ref{boundary}, where a cell $\mathcal{C}_1$ is joined to a cell $\mathcal{C}_2$ at a boundary $\mathcal{B}$. As the boundary has a non-zero velocity in the coordinates of the weak-field expansion (\ref{pnmetric}), it is not orthogonal to a surface of constant $t$ on either side. Instead, the two background spaces described by surfaces of constant $t$ intersect at $\mathcal{B}$, but do not overlap. This is required in order for the space-time to be smooth when the boundary is in motion \cite{Sanghai:2015wia}. It also explains why we do not have $\vec{n} \cdot \vec{\nabla} \phi=0$ at a reflection symmetric boundary, as the spatial derivatives in such an expression would be evaluated on spatial surfaces that do not overlap on either side of $\mathcal{B}$, and hence would not be directly comparable.

\section{General behaviour}\label{sec:gen}

In order to determine the equations describing the large-scale expansion of our model it will be useful to define averages over the volume of a cell, and over the surface area of a cell, as
\begin{equation}
\langle X \rangle_{\mathcal{C}} := \frac{1}{L^3} \int_{\mathcal{C}} X \mathrm{d}V \qquad {\rm and} \qquad \langle X \rangle_{\partial {\mathcal{C}}} := \frac{1}{6 L^2} \int_{\partial {\mathcal{C}}} X \mathrm{d}A \, ,
\end{equation}
where $\partial {\mathcal{C}}$ is boundary of the cell $\mathcal{C}$. We will also need Reynold's transport theorem:
\begin{equation}
\frac{\mathrm{d}}{\mathrm{d}t} \int_{\mathcal{C}} X \mathrm{d}V = \int_{\mathcal{C}} \frac{\partial X}{\partial t} \mathrm{d}V + \int_{\partial {\mathcal{C}}} X \boldsymbol{v} \cdot \boldsymbol{\mathrm{d}} \boldsymbol{A} \, ,
\end{equation}
where $\boldsymbol{v}$ is the velocity of the boundary of the cell, such that $\boldsymbol{v} \cdot \boldsymbol{\mathrm{d}} \boldsymbol{A} = \frac{1}{2} H L \, \mathrm{d}A$. Applying this to the average over a cell gives
\begin{equation} \label{com}
\frac{\mathrm{d}}{\mathrm{d}t} \langle X \rangle_{\mathcal{C}} = \langle \partial_t X \rangle_{\mathcal{C}} - 3 H \Delta \langle X \rangle 
\end{equation}
where we have defined $ \Delta \langle X \rangle \equiv \langle X \rangle_{\mathcal{C}} -\langle X \rangle_{\partial {\mathcal{C}}}$ (i.e. the difference between the average over the volume and the surface area of a cell). Finally, let us define the covariance of any two quantities $X$ and $Y$ as
\begin{equation} \label{cov}
{\rm Cov} (X,Y) \equiv \langle (X- \langle X \rangle_{\mathcal{C}}) (Y- \langle Y \rangle_{\mathcal{C}}) \rangle_{\mathcal{C}} = \langle X Y \rangle_{\mathcal{C}} - \langle X \rangle_{\mathcal{C}} \langle Y \rangle_{\mathcal{C}} \, ,
\end{equation}
and the variance as ${\rm Var} (X) \equiv {\rm Cov} (X,X)$. These are all of the ingredients we will need to determine the cosmological evolution of our construction.

First, we wish to determine the equivalent of the first Friedmann equation. This follows from the Gauss embedding equation, which we can write as
\begin{equation}
2 u^{\mu} u^{\nu} G_{\mu \nu} = {}^{(3)} \mathcal{R} + \tilde{K}^2 - \tilde{K}_{\mu \nu} \tilde{K}^{\mu \nu} \ ,
\end{equation}
where $u^{\mu} = (1, H x^i) + \mathcal{O}(\epsilon^2)$ is a congruence that is uniformly expanding, and comoving with the boundary. The quantities ${}^{(3)} \mathcal{R}$ and $\tilde{K}_{\mu \nu}$ are the Ricci curvature scalar and extrinsic curvature of a surface of constant $t$, respectively. Using this equation, together with the field equations of the theory, and integrating over a cell, gives\footnote{We have neglected the contribution of $(\vec{\nabla} \phi)^2$ to the energy density of the scalar field when calculating the equations for the large-scale expansion, in (\ref{em1}) and (\ref{em2}). This is justified by the post-Newtonian approach adopted from Sec. \ref{sec:pert} onwards, where gradient contributions to the stress-energy tensor would be higher-order corrections.}
\begin{equation} \label{em1}
\hspace{-2.5cm}
\boxed{
\left(\frac{\dot{a}}{a}\right)^2 = \frac{8 \pi G}{3} \langle C \rangle_{\mathcal{C}} \langle \rho \rangle_{{\mathcal{C}}} - \frac{\kappa}{a^2} + \frac{\Lambda}{3} +\frac{8 \pi G}{3} \left[ \frac{1}{2} \langle \Pi \rangle^2_{\mathcal{C}} + \langle V \rangle_{\mathcal{C}} \right]+ \frac{8 \pi G}{3} {\rm Cov}(C, \rho) + \frac{4 \pi G}{3} {\rm Var}(\Pi)} \, ,
\end{equation}
where we have made use of the definitions in Eq. (\ref{cov}), and where we have defined $\kappa \equiv \frac{1}{6} a^2 \langle {}^{(3)}R \rangle_{\mathcal{C}}$ and $\Pi \equiv {\partial_t \phi}$. The scale factor $a=a(t)$ should be understood to be proportional to the cell size $L(t)$, such that the Hubble rate can be defined as $H\equiv \dot{a}/a$. This equation is recognisable as a generalization of the first Friedmann equation (\ref{frw1}). Here it should be considered an emergent equation, from the post-Newtonian treatment of gravity within one of our cells.

Let us now derive the analogue of the second Friedmann equation (\ref{frw2}), in our inhomogeneous construction. The junction condition $K_{\mu \nu}=0$ implies that a point on the boundary can be taken to be following a geodesic \cite{Sanghai:2015wia}. This is a necessary condition for any reflection symmetric surface, which are all necessarily extremal, and therefore totally geodesic \cite{eis}. We therefore need to consider the equation of motion of a test particle on the boundary, which is given by (\ref{eq:motion}), and which to leading order is
\begin{eqnarray}
C(\phi)\,\frac{\dd \vec{v}}{\dd t}+ C'(\phi) \dot{\phi} \, \vec{v}
&= - C(\phi) \vec{\nabla} \Phi - C'(\phi) \vec{\nabla} \phi \, .
\end{eqnarray}
We interpret the second term on the left-hand side of this equation as a friction term, due to the time variation of the effective inertial mass $C(\phi)$ of test particles on the boundary. It can be seen from the junction condition on the scalar field (\ref{dphi}) that this term exactly cancels with the second term on the right-hand side. This would appear to have been hard to predict, as it corresponds to a perfect cancellation between the change in the particle's inertial mass and the fifth force due to the scalar field. These are two separate phenomena, which conspire to cancel due to the junction condition (\ref{dphi}), which seems remarkable.

After removing the cancelling terms, integrating over the boundary of the cell, and dividing by $C(\phi)$ and the cell's volume $\mathcal{V}$, we are left with
\begin{equation}
\frac{\ddot{a}}{a} = - \frac{1}{3 \mathcal{V}} \int_{\partial\mathcal{C}} \dd A \; \vec{n} \cdot \vec{\nabla}\Phi \, ,
\end{equation}
where we have again written $v=\frac{1}{2} H L$. Using Gauss's theorem on the right-hand side of this equation, and using the post-Newtonian equation for $\nabla^2 \Phi$ from Eq. (\ref{eq:field}), then gives
\begin{equation}
\frac{1}{\mathcal{V}}\int_{\partial\mathcal{C}} \dd A \; \vec{n} \cdot \vec{\nabla}\Phi
= \langle \nabla^2 \Phi \rangle_{\mathcal{C}}
= 4\pi G \left[ \langle{C(\phi) \rho} \rangle_{\mathcal{C}} + 2\langle \Pi^2 \rangle_{\mathcal{C}} - 2 \langle V(\phi)  \rangle_{\mathcal{C}} \right]- \Lambda \, ,
\end{equation}
which results in 
\begin{equation} \label{em2}
\hspace{-1.5cm}
\boxed{
\frac{\ddot{a}}{a} = -\frac{4 \pi G}{3} \langle C \rangle_{\mathcal{C}} \langle \rho \rangle_{{\mathcal{C}}} + \frac{\Lambda}{3}  -  \frac{8 \pi G}{3} \left[ \langle \Pi \rangle^2_{\mathcal{C}} -\langle V \rangle_{\mathcal{C}} \right]-\frac{4 \pi G}{3} {\rm Cov}(C,\rho) - \frac{8 \pi G}{3} {\rm Var}(\Pi) }\, .
\end{equation}
This is again strikingly similar to the second Friedmann equation (\ref{frw2}), but here has emerged from our patchwork construction of post-Newtonian cells.

It now remains to determine an emergent equation for the evolution of the scalar field $\phi$, which we can compare to the FLRW equation (\ref{frw3}). Taking the leading-order part of the Klein-Gordon equation (\ref{eq:scalar}) gives
\begin{equation}
-\ddot{\phi} + \nabla^2 \phi = V'(\phi) + C'(\phi) \rho  \, .
\end{equation}
In order to find the evolution of the large-scale part of $\phi$ we integrate the above equation over $\mathcal{C}$, and divide by its volume~$\mathcal{V}$, to find
\begin{equation}\label{eq:evolution_phibar}
\langle \ddot{\phi} \rangle_{\mathcal C} + \langle {V'(\phi)} \rangle_{\mathcal C} +   \langle C'(\phi) \rho \rangle_{\mathcal C}
= \frac{1}{\mathcal{V}} \int_{\partial\mathcal{C}} \dd A \; \vec{n} \cdot \vec{\nabla}\phi \ .
\end{equation}
The right-hand side of this equation can now be dealt with using the boundary conditions at $\mathcal{C}$, as specified in Eq. (\ref{dphi}):
\begin{equation} \label{ndphi}
\hspace{-1cm}
\frac{1}{\mathcal{V}} \int_{\partial\mathcal{C}} \dd A \; \vec{n} \cdot \vec{\nabla}\phi
= - \frac{1}{2 \mathcal{V}} \int_{\partial\mathcal{C}} \dd A \; H L \, \dot{\phi}
= - \frac{1}{2 L^3} \, 6L^2 \, H L \langle \dot{\phi} \rangle_{\partial {\mathcal C}}
= -3 H \langle \dot{\phi} \rangle_{\partial {\mathcal C}} \ ,
\end{equation}
where we have assumed that the cell is cubic. Commuting averaging and derivative operators using Eq.~(\ref{com}), and performing manipulations, then gives
\begin{equation} \label{em3}
\boxed{
\frac{\mathrm{d}}{\mathrm{d}t} \langle \Pi \rangle_{\mathcal{C}} + 3 H \langle \Pi \rangle_{\mathcal{C}} + \langle V' \rangle_{\mathcal{C}} + \langle C'\rangle_{\mathcal{C}} \langle \rho \rangle_{\mathcal{C}} = - {\rm Cov} (C',\rho)} \, ,
\end{equation}
where
$$
\frac{\mathrm{d}}{\mathrm{d}t} \langle \phi \rangle_{\mathcal{C}} = \langle \Pi \rangle_{\mathcal{C}} - 3 H \Delta \langle \phi \rangle \, .
$$
These equations can be seen to be very similar to the FLRW equation for the scalar field (\ref{eq:scalar}).

To complement the equations above, we can also derive evolution equations for the source terms in each of these equations, as follows:
\begin{eqnarray}
\frac{\mathrm{d}}{\mathrm{d}t} \langle \rho \rangle_{\mathcal{C}} &=& -3 H \langle \rho \rangle_{\mathcal{C}} \\
\frac{\mathrm{d}}{\mathrm{d}t} \langle C \rangle_{\mathcal{C}} &=& \langle C' \rangle_{\mathcal{C}} \langle\Pi \rangle_{\mathcal{C}} + {\rm Cov}(C',\Pi) - 3 H \Delta \langle C \rangle \\
\frac{\mathrm{d}}{\mathrm{d}t} \langle V \rangle_{\mathcal{C}} &=& \langle V' \rangle_{\mathcal{C}} \langle\Pi \rangle_{\mathcal{C}} + {\rm Cov}(V',\Pi) - 3 H \Delta \langle V \rangle \, .
\end{eqnarray}
The Var, Cov and $\Delta$ terms in the equations derived in this section are not present in the homogeneous and isotropic FLRW cosmologies of these theories. They are back-reaction terms, which can change the large-scale evolution that emerges in a universe containing non-linear structures. In what follows, we will evaluate them in the case where the scalar field can be expanded perturbatively, as well as when a non-linear screening mechanism occurs.

\section{Perturbed scalar fields} \label{sec:pert}

In many cases it is a good approximation to expand the scalar field $\phi$ around a background, such that
\begin{equation} \label{phipert}
\phi = \bar{\phi}(t) + \delta \phi + \mathcal{O}(\epsilon^4) \, ,
\end{equation}
where $\bar{\phi}(t) \sim \epsilon^0$ is the homogeneous background value, and $\delta \phi \sim \epsilon^2$ is the inhomogeneous perturbation around that background, which carries the fifth force. These orders of smallness are chosen so that leading-order contribution from $\phi$ appears in the leading-order parts of the field equations (\ref{eq:field}) and (\ref{eq:scalar}).

In this section we will choose to identify the background part of $\phi$ with the spatial average, so that
\begin{equation}
\langle \phi \rangle_{\mathcal{C}} = \bar{\phi} + \mathcal{O}(\epsilon^4) \qquad {\rm or, \,\, equivalently,} \qquad \langle \delta \phi \rangle_{\mathcal{C}} =0 + \mathcal{O}(\epsilon^4) \, .
\end{equation}
While the background value of a scalar field does not normally need to be specified in this way, we will find that in the present case it is essential. Failure to do so would mean that we would not recover the FLRW equations (\ref{frw1})--(\ref{frw3}).

Let us start by considering the emergent equations (\ref{em1}) and (\ref{em2}) from above, which are analogues of the first and second Friedmann equations in our inhomogeneous construction. In these equations the conformal factor $C(\phi)$ occurs in terms that multiply the energy density $\rho$. As $\rho \sim \epsilon^2$ in the post-Newtonian expansion, we only require the leading order contribution to $C(\phi)$, which is simply $C(\bar{\phi})$. This term is clearly homogeneous, which means that to the order we require we have
\begin{equation}
\langle C(\phi) \rangle_{\mathcal{C}} = C(\bar{\phi}) \qquad {\rm and} \qquad {\rm Cov} (C,\rho) =0 \, .
\end{equation}
Now, the terms that contain a factor $\langle V(\phi) \rangle_{\mathcal{C}}$ are only required up to order $\epsilon^2$, which means that we can perform a Taylor expansion to find
\begin{equation}
\langle V(\phi) \rangle_{\mathcal{C}} = V(\bar{\phi}) + V'(\bar{\phi}) \langle \delta\phi \rangle_{\mathcal{C}} + \mathcal{O}(\epsilon^4)
= V(\bar{\phi})+ \mathcal{O}(\epsilon^4) \, ,
\end{equation}
where in the last equality we have used the result that $\langle \delta\phi \rangle_{\mathcal{C}}=0$. Finally, equations (\ref{em1}) and (\ref{em2}) contain terms involving $\Pi$. As time derivatives add orders of smallness in the post-Newtonian formalism, we have to the required order that
\begin{equation}
\langle \Pi \rangle_{\mathcal{C}} = \dot{\bar{\phi}}(t) + \mathcal{O}(\epsilon^3) \qquad {\rm and} \qquad {\rm Var} (\Pi) = 0+ \mathcal{O} (\epsilon^4) \, ,
\end{equation}
where the second result derives from the leading-order part of $\Pi$ being homogeneous.

Substituting the results above into the emergent equations (\ref{em1}) and (\ref{em2}) gives that, when the scalar field can be expanded as in Eq. (\ref{phipert}), the analogue of the first Friedmann equation becomes 
\begin{equation}
\hspace{-0.75cm}
\boxed{
\left( \frac{\dot{a}}{a} \right)^2
= \frac{8 \pi G C(\bar{\phi})}{3} \langle {\rho} \rangle_{\mathcal{C}}
	- \frac{\kappa}{a^2}
	+ \frac{\Lambda}{3}
	+ \frac{8\pi G}{3}
		\left[ \frac{1}{2} \dot{\bar{\phi}}^2 + V(\bar{\phi}) \right]} \, ,
\end{equation}
while the second becomes
\begin{equation}
\boxed{
\frac{\ddot{a}}{a} = -\frac{4\pi G C(\bar{\phi})}{3} \langle {\rho} \rangle_{\mathcal{C}} + \frac{\Lambda}{3}- \frac{8\pi G}{3}\left[\dot{\bar{\phi}}^2 - V(\bar{\phi})\right]}\, .
\end{equation}
These two equations are plainly identical in form to the FLRW equations (\ref{frw1}) and (\ref{frw2}), with $p=0$ and with the Friedmann energy density and scalar field values being replaced by the spatial average $\langle {\rho} \rangle_{\mathcal{C}}$ and the background value $\bar{\phi}$. Let us now consider the emergent equation for the scalar (\ref{em3}).

In equation (\ref{em3}) we have a $\langle V' \rangle_{\mathcal{C}}$, and terms containing $C'$ multiplied by the energy density~$\rho$. These terms can be dealt with in exactly the same way as those that contained $\langle V \rangle_{\mathcal{C}}$ and $C$ in Eqs. (\ref{em1}) and (\ref{em2}), just with an extra derivative with respect to $\phi$ in each case. This means
\begin{equation}
\hspace{-1cm} 
\langle V' \rangle_{\mathcal{C}} = V'(\bar{\phi}) + \mathcal{O}(\epsilon^4) \, , \qquad
\langle C' \rangle_{\mathcal{C}} = C'(\bar{\phi}) + \mathcal{O} (\epsilon^2) \qquad
{\rm and} \qquad {\rm Cov}(C',\rho) =0 \, .
\end{equation}
The only other terms that need to be considered are the time derivative of the average of $\phi$, and the term corresponding to the difference in averages of $\phi$ over the volume and the boundary of the cell. As the leading-order part of the average of $\phi$ is homogeneous, these are simply
\begin{equation}
\frac{\dd}{\dd t} \langle \phi \rangle_{\mathcal{C}} = \frac{\dd \bar{\phi}}{\dd t} + \mathcal{O}(\epsilon^3) \qquad {\rm and} \qquad
\Delta \langle \phi \rangle = 0 + \mathcal{O}(\epsilon^2) \, .
\end{equation}
Substituting these results into the emergent scalar field equation (\ref{em3}), and eliminating $\Pi$, gives
\begin{equation}\label{eq:evolution_phibar_result}
\boxed{
\ddot{\bar{\phi}} + 3\frac{\dot{a}}{a} \dot{\bar{\phi}} + V'(\bar{\phi}) +  C'(\bar{\phi}) \langle {\rho} \rangle_{\mathcal{C}}
= 0 } \, ,
\end{equation}
which is again plainly identical in form to the FLRW equation (\ref{frw3}), with $p=0$ and the relevant quantities replaced by their averages and background values. We therefore have a clear emergence of the expected FLRW behaviour in situations in which the scalar field can be expanded as in (\ref{phipert}), provided we identify the values of $\phi$ and $\rho$ appropriately. {  We note that the emergence of FRLW behaviour in the Einstein frame also guarantees the corresponding emergence of FLRW behaviour in the Jordan frame.}

\section{Non-perturbative screening mechanisms} \label{sec:screen}

In the preceding section we allowed the scalar field to be perturbatively expanded around a background value, $\bar{\phi}$. While we expect such an approach to be valid in a wide range of applications of scalar-tensor theories, there is also a substantial body of work in which non-perturbative ``screening'' mechanisms play a role \cite{vain, cham, kmo, symm}. In this section we wish to study if and how such non-linear mechanisms affect the emergent large-scale cosmological expansion.

The particular theory we wish to consider, as an illustrative example, is the one specified by
\begin{equation}
V = \frac{V_0}{\phi} \qquad {\rm and} \qquad C = \exp (\phi/M) \, ,
\end{equation}
where $V_0$ and $M$ are constants. This theory is known to exhibit the ``Chameleon'' screening mechanism \cite{cham}, which has been particularly well-studied in the literature \cite{crev}. Throughout this section we make the usual assumption that $\phi/M \ll 1$, such that $C \simeq 1$ and $C' \simeq M^{-1}$. {  This means that the geometry of space-time will be similar in both the Einstein and Jordan frames, as $\tilde{g}_{\mu \nu} \simeq g_{\mu \nu}$, while also allowing us to write the Einstein-frame scalar field equation as}
\begin{equation}\label{screened}
\nabla^2 \phi  \simeq -\frac{V_0}{\phi^2} + \frac{\rho}{M} + \ddot{\phi} \, .
\end{equation}
The boundary condition (\ref{dphi}) then allows us to solve this equation for any given distribution of mass $\rho$. {  For this we make use of the SELCIE code, developed in Ref. \cite{selc}, which has since been modified to incorporate Neumann boundary conditions}.

In order to gain intuition on these theories, we will consider specific realizations of inhomogeneous matter distributions within our cubic lattice cells. While the SELCIE code is able to deal with arbitrary distributions of matter, here we restrict ourselves to the simplest possible configuration: a spherical body, with evenly distributed mass, at the centre of the cell. This is done so that we can build intuition on the aspects of the screening mechanism that are relevant for our problem, without having to concern ourselves with the more complex behaviour that results from more sophisticated distributions of matter \cite{shapes}. 

\begin{figure}
\center
\includegraphics[width=18cm]{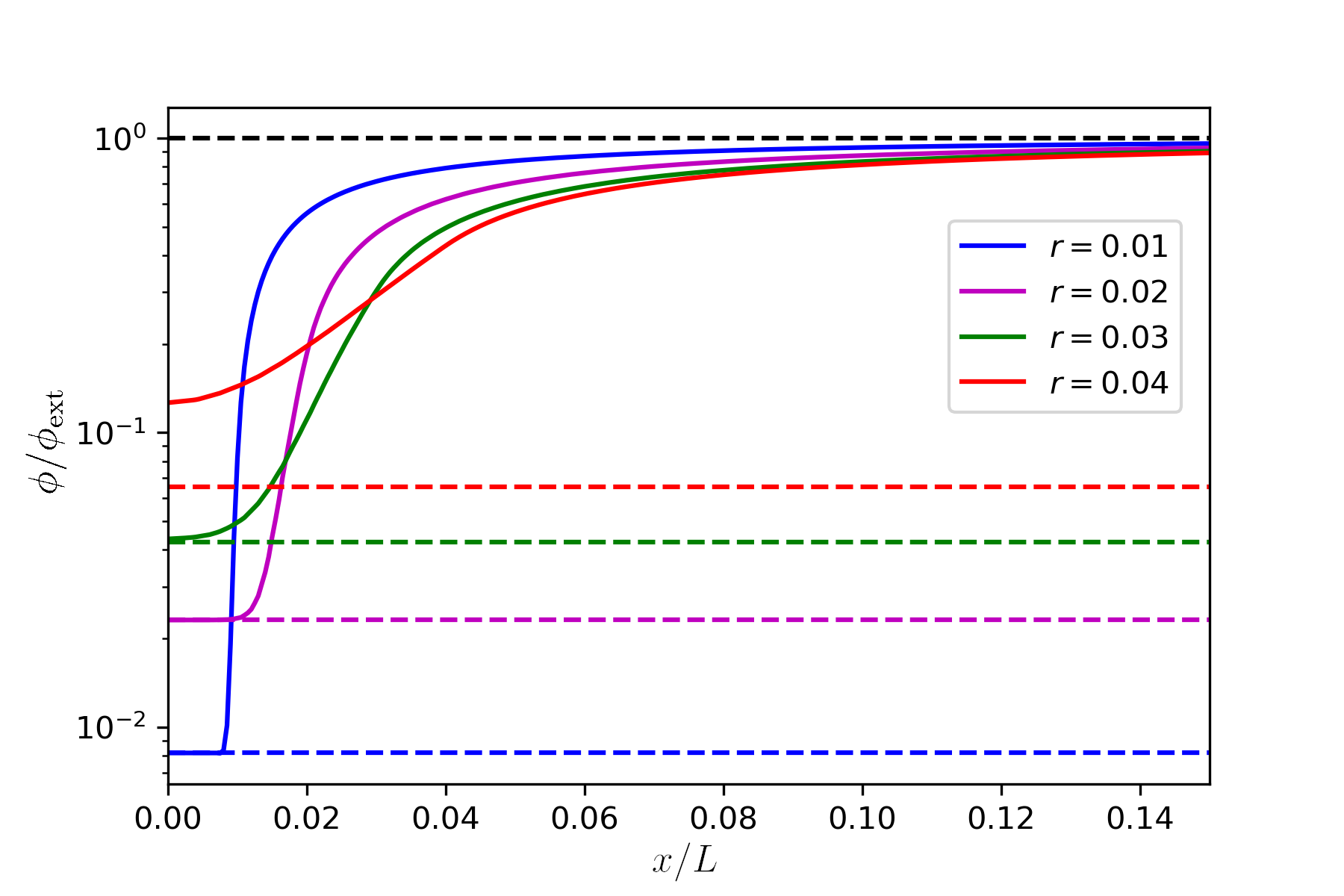}
\caption{Scalar field profiles along a line connecting the centre of a cell and the centre of a cell face. Curves correspond to spheres of radius $0.04$ (red), $0.03$ (green), $0.02$ (purple) and $0.01$ (blue), and range from completely unscreened (red) to strongly screened (blue), with the screened value shown in Eq. (\ref{phiint}) indicated by dashed horizontal lines. Distances are given in units of the cell length, $L$, and the scalar field is in units of $\phi_{\rm ext}$ (from Eq. (\ref{phiex}), and as indicated by the horizontal dashed black line).}  \label{spheres}
\end{figure}

The results of our simulations are shown in Fig. \ref{spheres}, where spheres of different sizes (but equal mass) are considered. In this figure the field is defined in units of $\phi_{\rm ext}$, which is the field value in the region far from the matter source. Using Eq. (\ref{screened}) this is
\begin{equation} \label{phiex}
\phi_{\rm ext} \simeq \sqrt{\frac{V_0}{\ddot{\phi}}} \, ,
\end{equation}
where in this region we have assumed that $\rho$ is negligible (i.e. we are close to vacuum outside of the body). In the presence of strong screening, $\rho$ dominates over $\ddot{\phi}$ and so the effective potential is minimised when the first two terms on the right-hand side of Eq. (\ref{screened}) cancel. This happens when the scalar field in the interior of the body takes the value
\begin{equation} \label{phiint}
\phi_{\rm int} \simeq \sqrt{\frac{M V_0}{\rho}} \, ,
\end{equation}
and is indicated in Fig. \ref{spheres}, for each source, by a horizontal dashed line. We would not normally expect the time-dependence of $\phi$ to have a noticeable effect on this behaviour, in non-linear, gravitationally-bound structures where $\rho$ does not change rapidly with time, and where cosmological dynamics are largely inconsequential for the local gravitational physics. In this case we have a solution where $\phi$ in the interior region is approximately constant in space and time. Eqs. (\ref{phiint}) and (\ref{phiex}) then give the two field values that the screened curves in Fig. \ref{spheres} approaches in the regions interior and exterior to the screened body. The value of $\phi$ jumps between these values at the edge of the body, and in the exterior we can treat the value of $\phi$ as being given by a perturbative expansion about the value given in Eq. (\ref{phiex}).

The red line in Fig. \ref{spheres} corresponds to a totally unscreened mass. In this case we have verified that $\langle V'(\phi) \rangle \simeq V'(\langle \phi \rangle )$ to a high degree of accuracy, which demonstrates the applicability of the perturbative approach from the previous section to unscreened masses in this theory. In cosmologies in which all mass is unscreened, we therefore expect the large-scale expansion to be well-represented by the FLRW solutions of the theory. In contrast, if one considers smaller and smaller spheres, with higher and higher densities, this result no longer holds. This is shown in Fig. \ref{spheres} by the curves with the lowest central values of $\phi$, in which screening is occurring, and in which non-perturbative behaviour is being displayed. We have verified that in such cases $\langle V'(\phi) \rangle \neq V'(\langle \phi \rangle )$, and as such the treatment in the previous section does not apply. It is these cases that we will focus on in the remainder of this section, in order to determine the cosmological effects of screening.

\begin{figure}
\center
\includegraphics[width=18cm]{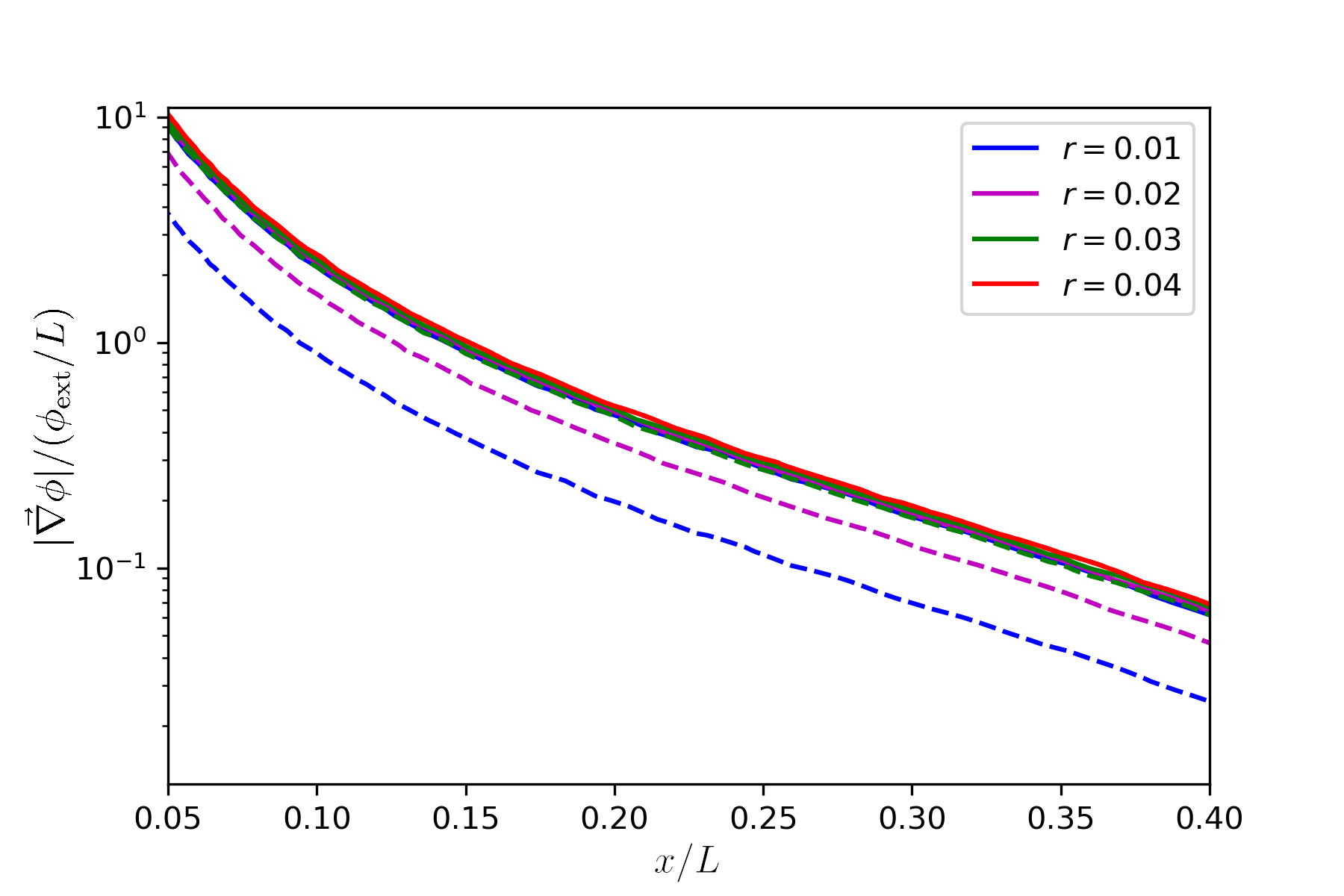}
\caption{The magnitude of the gradient of the scalar field, outside of the spherical body at the centre of the cell. Colours correspond to the configurations from Fig. \ref{spheres}, with distances in units of the cell length $L$, and $\phi_{\rm ext}$ as defined in Eq. (\ref{phiex}). Dashed lines correspond to raw values, and solid lines (stacked very closely together) to values rescaled by the ratio $m/m_{\rm shell}$.}  \label{dphifig}
\end{figure}

In order to derive emergent cosmological equations we need to average each of the terms in Eq. (\ref{screened}) over the volume of the cell. Integrating the $\nabla^2 \phi$ term over the cell gives us the following integrated scalar fifth-force on the boundary:
\begin{equation} \label{sforce}
\int_{\mathcal{C}} \nabla^2 \phi \, \dd V =\int_{\partial \mathcal{C}} \vec{\nabla} \phi \cdot \dd \vec{A} \, .
\end{equation}
In the space exterior to the body we expect to be able to use perturbation theory to obtain
\begin{equation} \label{sex}
\nabla^2 \phi \simeq -\frac{V_0}{\bar{\phi}^2} +   \frac{2 V_0}{\bar{\phi}^3} \, \delta \phi +\ddot{\bar{\phi}} \, , \hspace{2cm} {\rm ( in\; the \; exterior \; of\; the \; screened \; body )}
\end{equation}
where we have defined $\bar{\phi} \equiv \langle \phi \rangle_{\rm ext}$ and $\delta \phi \equiv \langle \phi \rangle_{\rm ext}- \bar{\phi}$, and where $\langle X \rangle_{\rm ext}$ means the spatial average of $X$ over the region exterior to the screened mass. We have assumed that $\rho=0$ in the exterior region, and neglected $\delta \ddot{\phi}$, as such a term is at higher order in our post-Newtonian expansion. In the region of space interior to the body the potential $V(\phi)$ is expected to cancel the mass density at all points, except within a thin shell near its surface \cite{cham}. Within this region the density dominates over the potential term in Eq. (\ref{screened}), allowing it to be approximated as
\begin{equation} \label{sin}
\nabla^2 \phi \simeq \frac{\rho_{\rm shell}}{M} \, ,\hspace{4.75cm} {\rm (at\; the\; surface \; of \; the \; screened \; body)} 
\end{equation}
where $\rho_{\rm shell}$ is the unscreened energy density inside the shell, and we have assumed $\phi$ is not changing in time within the body. Using (\ref{sex}) and (\ref{sin}) to evaluate the left-hand side of (\ref{sforce}), and the boundary condition (\ref{dphi}) for the right-hand side, we get
\begin{equation}\label{phiscreened}
\boxed{
\alpha \, \ddot{\bar{\phi}} +3 \frac{\dot{a}}{a} \, \dot{\bar{\phi}} +\alpha \,  V'(\bar{\phi})  + \frac{\langle \rho_{\rm shell} \rangle_{\mathcal{C}}}{M} =0 }\, ,
\end{equation}
where $\alpha\equiv\mathcal{V}_{\rm ext}/\mathcal{V}$ is the volume of the exterior region $\mathcal{V}_{\rm ext}$, as a fraction of the total spatial volume of the cell $\mathcal{V}$.
The validity of this approach is demonstrated using our numerical simulations in Fig. \ref{dphifig}, where the scalar field profile outside of the central body is the same independent of the presence or absence of screening, up to a multiplicative factor that is accurately modelled by the ratio of the thin shell mass to the total mass.

Using the general equations for the analogue of the first and second Friedmann equations, (\ref{em1}) and (\ref{em2}), we find for the present case that 
\begin{equation}\label{f1screened}
\boxed{
\left( \frac{\dot{a}}{a} \right)^2 =  
\frac{8\pi G}{3} \left[\langle \rho \rangle_{\mathcal{C}}+ \alpha \left( \frac{\dot{\bar{\phi}}^2}{2} +V(\bar{\phi}) \right) +(1-\alpha) \, \sqrt{ \frac{\rho V_0}{M}} \right]
  - \frac{\kappa}{a^2} +\frac{\Lambda}{3} }\, ,
\end{equation}
and
\begin{equation}\label{f2screened}
\boxed{
\frac{\ddot{a}}{a}
=-\frac{8\pi G}{3} \left[ \frac{1}{2}\langle \rho \rangle_{\mathcal{C}} + \alpha \left(\dot{\bar{\phi}}^2 -V(\bar{\phi}) \right)-(1-\alpha) \sqrt{ \frac{\rho V_0}{M}} \right] + \frac{\Lambda}{3} 
}\, ,
\end{equation}
where we have neglected the contribution from within the thin shell to the total energy density and pressure of the scalar field inside the cell. We note that while it was convenient to use the general equations (\ref{em1}) and (\ref{em2}) to find the above, we chose not do to use Eq. (\ref{em3}) to derive the scalar field equation (\ref{phiscreened}). This is because Eq. (\ref{em3}) gives an evolution equation for the average of $\phi$ throughout the cell, $\langle \phi \rangle_{\mathcal{C}}$, while for our current purposes it is convenient to derive an equation for the average of $\phi$ in the unscreened region only.

In the limit where the fractional volume occupied by the screened mass is negligibly small, as one might expect in a universe in which all matter has collapsed into dense objects, we will have $\alpha \rightarrow 1$. In such a case the equations above reduce to the Friedmann equations (\ref{frw1})--(\ref{frw3}), with the mass density in the scalar field equation being replaced by the density of the thin shell. The scalar component of our gravitational theories will then only be apparent through the contribution of this one term, which is generically expected to be small when the shell is thin, so that the cosmology will effectively behave as if it were governed by general relativity in the presence of a minimally coupled scalar field. When the value of $\alpha$ differs from one, we can see that the energy density of the scalar field $\rho_{\phi}$, and the gravitational energy density $\rho_{\phi}+3p_{\phi}$, contribute with a multiplicative factor of $\alpha$ in the Friedmann equations (\ref{f1screened}) and (\ref{f2screened}). This shows that the scalar field in the screened region is not contributing to these terms, but is instead accounted for by the term proportional to $(1-\alpha)$. In the emergent scalar field equation (\ref{phiscreened}), the factors of $\alpha$ act to reduce the contribution of the terms that correspond to the derivative of the scalar field energy density.

\section{Discussion} \label{sec:disc}

We have considered post-Newtonian cosmological model building in the context of scalar-tensor theories of gravity. We have shown that when the scalar field can be treated perturbatively, that the emergent Friedmann equations that govern the large-scale cosmological expansion are identical to those of homogeneous and isotropic cosmological models. This indicates that the presence of non-linear structures does not have a strong back-reaction effect in such cases. In contrast, in the presence of the non-perturbative chameleon screening mechanism the emergent cosmological behaviour differs from the Friedmann solutions of the theory, showing that strong back-reaction can occur when non-linear, screened structures start appearing in the Universe. In particular, screened mass drops out of the Klein-Gordon equation for the scalar field. 

This behaviour is qualitatively different from that which was obtained for $f(R)$ models in Refs. \cite{dun1, dun2, cano}, where it was shown that cosmological evolution and the local Newtonian limit can be incompatible even in the absence of screening\footnote{This inequivalence is true despite the close relationship between $f(R)$ and scalar-tensor theories \cite{review2}, and is due to the existence of an algebraic relationship between the Ricci scalar of space-time, $R$, and the scalar degree of freedom of the theory in $f(R)$ gravity.}. {  We expect our results to be useful for cosmological simulations in which structure forms as the Universe evolves, and in which screening develops at late times. For any non-negligible amount of screened mass, we find that there will be consequences for the evolution of the background cosmology, which we consequently expect to have an influence on observational probes. These include direct consequences for probes of the background expansion itself (such as Hubble diagrams and baryon accoustic oscillations), as well as indirect consequences for other obervables that are sensitive to the expansion of the Universe (such as the growth rate of structure \cite{baker} and observations of the cosmic microwave background \cite{thomas}).

Our results are derived for the simple scalar-tensor theories with a canonical kinetic term and a universal non-minimal coupling between the scalar and tensor degrees of freedom, as given by the action in equation (\ref{action}). These are the theories that are most usually studied in the context of the chameleon screening mechanism \cite{cham}, but they are by no means the most general theories that could be considered. More general theories are discussed in the reviews \cite{review1, review2}, as well as the very many references within them, to which we refer the reader for further details. It would be an interesting task to study the degree to which the behaviour discovered in the present work applies to these more general theories, as well as the extent to which similar conclusions might be drawn for other types of screening mechanism \cite{vain, kmo}. We leave this work for future studies.}

\section*{Acknowledgements}

We thank Lorenzo Reverberi for discussions at early stages of this project. CB and TC acknowledge support from STFC grant ST/P000592/1. PF received the support of a fellowship from “la Caixa” Foundation (ID 100010434) at the early stages of this project. The fellowship code is LCF/BQ/PI19/11690018.

\end{document}